\documentclass[prl,reprint,amsfonts,amsmath,amssymb,showpacs,twocolumn]{revtex4}

\usepackage[dvipdfmx]{graphicx}
\usepackage{dcolumn}
\usepackage{bm}
\usepackage{amsthm}
\usepackage{mathrsfs}
\usepackage{comment}
\usepackage{color}
\begin{document}
\title{Topological Influence between Monopoles and Vortices: \\ a Possible Resolution of the Monopole Problem}

\author{Shingo Kobayashi$^1$}
\author{Michikazu Kobayashi$^2$}
\author{Yuki Kawaguchi$^1$}

\author{Muneto Nitta$^3$}
\author{Masahito Ueda$^{1,4}$}
\affiliation{$^1$Department of Physics, University of Tokyo, 7-3-1, Hongo, Bunkyo-ku, Tokyo 113-0033, Japan \\  $^2$Department of Basic Science, University of Tokyo, 3-8-1, Komaba, Meguro-ku, Tokyo, 153-8902, Japan \\ $^3$Department of Physics, and Research and Education Center for Natural Sciences, Keio University, 4-1-1 Hiyoshi, Yokohama, Kanagawa 223-8511, Japan
 \\ $^4$ERATO Macroscopic Quantum Control Project, JST, 2-11-16, Yayoi, Bunkyo-ku, Tokyo, 113-8656, Japan}
\date{\today}
\begin{abstract}
 Grand unified theories of fundamental forces predict that magnetic monopoles are inevitable in the Universe because the second homotopy group of the order parameter manifold is $\mathbb{Z}$. We point out that monopoles can annihilate in pairs due to an influence of Alice strings. As a consequence, a monopole charge is charactarized by $\mathbb{Z}_2$ rather than $\mathbb{Z}$ if the Universe can accommodate Alice strings, which is the case of certain grand unified theories. 
\end{abstract}
\maketitle

Spontaneous symmetry breaking plays a pivotal role in many subfields of physics
from condensed matter to high energy physics and cosmology.
When the full symmetry $G$ is spontaneously broken to its subgroup $H$,
the order parameter manifold is $M = G/H$.
Monopoles may exist when the order parameter manifold $M$
has a nontrivial second homotopy group $ \pi_2(M)$.
Grand unified theories (GUTs) with simple group $G$ were proposed
to unify three fundamental forces in Nature:
electromagnetic, weak and strong interactions. 
Since the unbroken subgroup $H$ of the unified group $G$ of GUTs
must contain the electromagnetic group $U(1)$,
one can show $\pi_2(G/H) \cong \pi_1(H) \cong \mathbb{Z}$ \cite{Preskill:1979zi}.
This is a notorious cosmological problem -- the monopole problem;
the magnetic monopoles are inevitable in the Universe
in contradiction to current observations \cite{Zeldovich:1978wj,Preskill:1979zi}.
In fact, it has provided one of the primary motivations of inflationary universe;
inflation can sweep out monopoles out of our horizon.
 It has widely been believed that $\pi_2(G/H) \cong \mathbb{Z} $ is
the necessary and sufficient topological condition for the existence of monopoles.
In this Letter, we point out a loophole in this theorem
if the theory admits Alice strings \cite{Schwarz;1982, Vilenkin;1994}, which is the case of certain GUTs based on SU(5) or SO(10) group.

An Alice string is typically characterized by $\pi_1 (S^2 /\mathbb{Z}_2)$, which has only one nontrivial equivalence class. Mermin showed that a monopole changes its sign while it rotates around an Alice string \cite{Mermin;1979,Trebin;1982,Michel;1980,Mineev;1998,Volovic;2003}. However this presents a difficulty in the definition of the monopole charge because it seems to be inconsistent with $\pi_2 (G/H) \cong \mathbb{Z}$. In this Letter, we resolve this problem by showing that monopoles and Alice strings should as a whole be described by a different homotopy group $\mathbb{Z}_2 \times \mathbb{Z}_2$ even though $\pi_2(G/H) = \mathbb{Z}$. This finding leads to a surprising result: even if there exist no Alice strings in the Universe, monopoles can annihilate pairwise through spontaneous 
pair creation of Alice strings, where an even number of Alice strings are topologically equivalent to no strings or an Alice string loop. This result imposes severe restrictions on the existence of monoples in the Universe; there are only two cases in which there is a single monopole or no monopole at all.

\begin{figure}[htbp]
\centering
\includegraphics[width=5cm ]{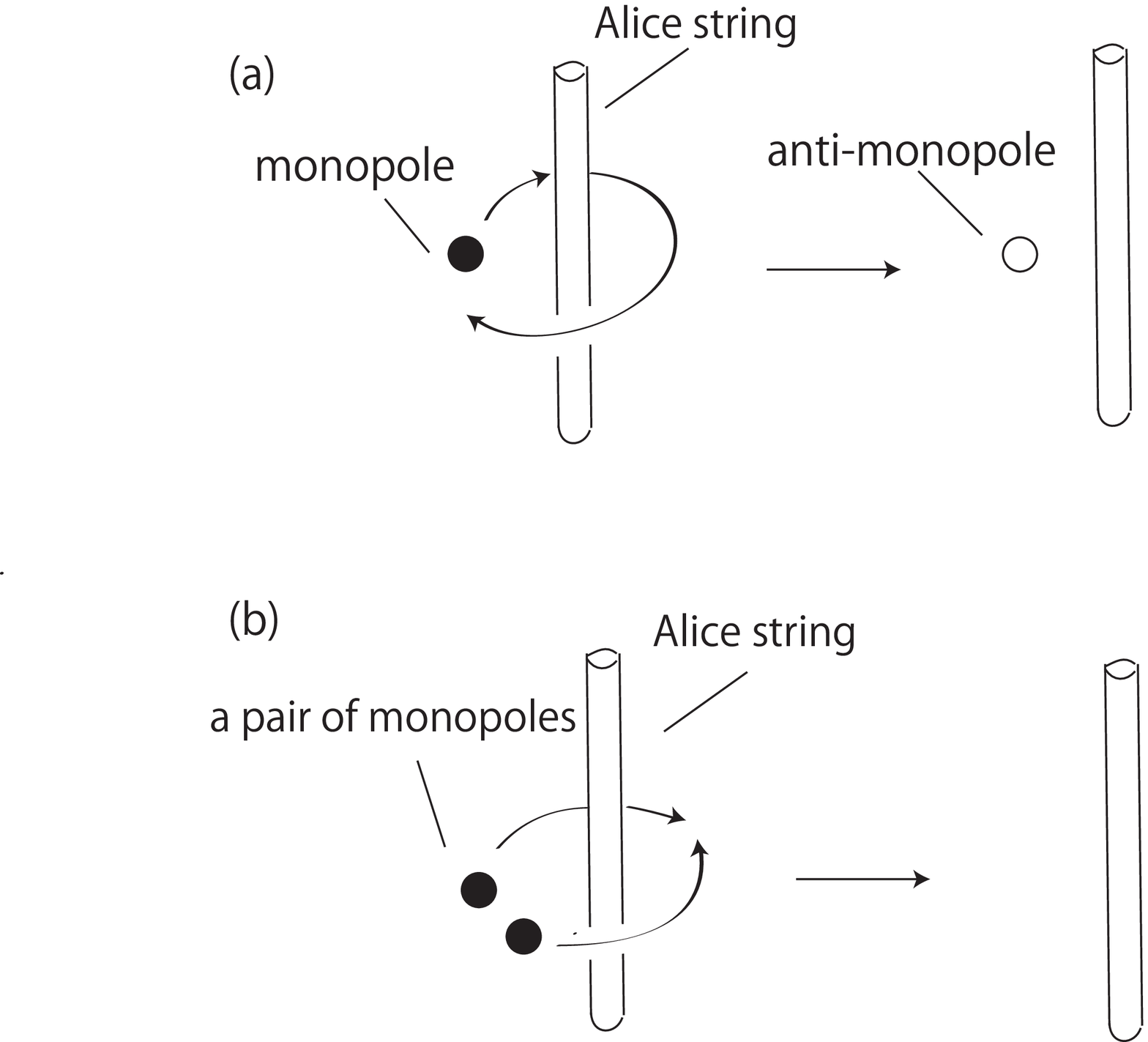}
\caption{(a) A monopole can be continuously transformed into an anti-monopole by rotating along a loop enclosing an Alice string. (b) When a pair of monopoles having the same charge goes past an Alice string on different sides (left figure) or equivalently, when an Alice string passes through a pair of monopoles, one monopole changes its sign relative to the other. As a consequence, monopoles can annihilate in pairs (right figure). }\label{letter-0}
\end{figure}

An influence of vortices on monopoles is represented by the commutator subgroup between $\pi_1 (M)$ and $\pi_2 (M)$. This commutator subgroup provides a key ingredient to find whether there is an influence among topological defects. When $\pi_2 (M) $ does not commute with $\pi_1 (M)$, $\pi_2 (M)$ alone cannot determine stable topological charges. This is the case when the order parameter manifold $M$ accommodates an Alice string, as shown in Fig. \ref{letter-0} (a), where 
 a monopole can be continuously transformed into an anti-monopole by rotating along a loop enclosing an Alice string. Such a phenomenon has been known as an influence of a disclination on a point defect in the nematic phase of liquid crystal \cite{Mermin;1979,Trebin;1982,Michel;1980,Mineev;1998,Volovic;2003}. In this Letter, we point out that, instead of the ordinary homotopy groups $\pi_2 (M)$, we should use a different homotopy group, namely the second Abe homotopy group $\kappa_2 (M)$ \cite{Abe;1940}, and that Alice strings and monopoles as a whole belong to a nontrivial homotopy group due to their mutual influence  \cite{footnote-1}.
 We apply it to classify monopoles that coexist with vortices. The corresponding $\kappa_2 (M)$ is isomorphic to a semidirect product of $\pi_1 (M)$ and $\pi_2 (M)$:
\begin{equation}
 \kappa_2 (M) \cong  \pi_1 (M) \ltimes \pi_2 (M),
\end{equation}
where the semidirect product $\ltimes$ means that for given $\alpha,\alpha' \in \pi_2 (M) $ and $ \gamma,\gamma' \in \pi_1 (M)$, a multiplication of the semidirect product is defined by 
\begin{equation}
\begin{split}
 (\gamma, \alpha) \ast (\gamma' , \alpha' )& = (\gamma \ast \gamma' , [\gamma'^{-1}, \alpha ] \ast \alpha \ast \alpha') \\
 [\gamma'^{-1},\alpha ] & \equiv \gamma'^{-1} \ast \alpha \ast \gamma' \ast \alpha^{-1}, \label{Abe-0}
 \end{split}
\end{equation}
where $\ast$ denotes a loop product \cite{Nakahara;2003,G.Whitehead;1978}.
  We can show that the commutator subgroup between $\pi_1 (M)$ and $\pi_2 (M)$ is included in the group structure of $\kappa_2 (M)$. If $\pi_2 (M)$ commutes with $\pi_1 (M)$, the multiplication law becomes $(\gamma, \alpha) \ast (\gamma' , \alpha' ) = (\gamma \ast \gamma' ,   \alpha \ast \alpha')$, which indicates that $\kappa_2 (M)$ is isomorphic to the direct product of $\pi_1 (M)$ and $\pi_2 (M)$. When we consider $M = S^2 /\mathbb{Z}_2$, the commutation relation becomes $[\gamma'^{-1},\alpha] = \alpha^{-2}$ \cite{Eilenberg;1939}. Therefore, $\pi_1(S^2/\mathbb{Z}_2)$ and $\pi_2 (S^2 /\mathbb{Z}_2)$ do not commute. As shown later, the second Abe homotopy group for $S^2 /\mathbb{Z}_2$ is isomorphic to $ \mathbb{Z}_2 \ltimes \mathbb{Z}$. In real situations, however, we have to take a conjugacy class of the second Abe homotopy group \cite{footnote-2}. 
 Therefore, we finally obtain topological charges as 
\begin{equation}
 \kappa_2 (S^2/\mathbb{Z}_2) / \sim \ \  \cong \mathbb{Z}_2 \times \mathbb{Z}_2, \label{Abe-1}
\end{equation}
where $\sim$ means that we take a conjugacy class of $\pi_2 (M) $ in terms of $\pi_1 (M)$ due to the base point dependence. The first $\mathbb{Z}_2$ and second $\mathbb{Z}_2$ on the right-hand side of Eq. (\ref{Abe-1}) represent Alice strings and monopoles, respectively. As a consequence, for $M= S^2 / \mathbb{Z}_2$, the monopole charge is always charactarized by $\mathbb{Z}_2$. Thus, the monopole problem does not occur if the order parameter manifold is $S^2 /\mathbb{Z}_2$. 
This result is applicable to a nematic phase in liquid crystals \cite{Mermin;1979,Trebin;1982,Michel;1980,Mineev;1998,Volovic;2003} and the polar phase in a spin-1 Bose-Einstein condensate \cite{Leonhardt;2000} because the corresponding order parameter manifolds include $S^2 /\mathbb{Z}_2$. Therefore, the pair  annihilation of monopoles due to the influence of Alice strings can be observed in the laboratory.
We will also discuss an entanglement among higher homotopy groups, which implies hitherto unknown topological excitations. 

\begin{figure}[htbp]
\centering
\includegraphics[width=5cm]{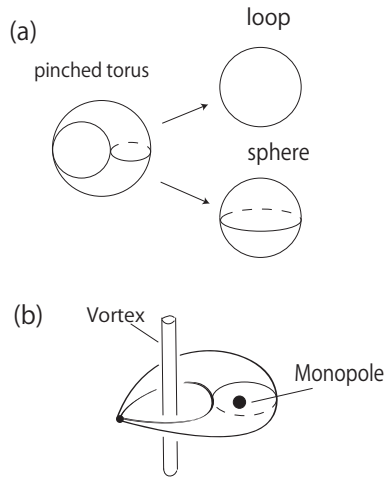}
\caption{(a) A pinched torus reduces to a loop if we shrink the torus to a line, or to a sphere  if we shrink a hole to a point. (b) The second Abe homotopy is described by a pinched torus that can accomodate an Alice string and a monopole simultaneously. By using the Abe homotopy group, the nontrivial influence between the first and second homotopy groups can be described.}\label{letter-1}
\end{figure}

We first prove Eq. (\ref{Abe-1}). The fundamental group $\pi_1 (M)$ is defined by a set of mappings from $S^1$ to $M$, and the second homotopy group $\pi_2$ is definded by a set of mappings from $S^2$ to $M$. Let map $f$ be a reference element of $\gamma \in \pi_1 (M)$. If $f$ is a mapping from a loop surrounding a line defect to an order parameter manifold, it cannot continuously shrink to a point in the manifold. Thus, $f $ represents a nontrivial vortex. Likewise, if map $g $, which is a reference element of $\alpha \in \pi_2 (M)$, is a mapping from a sphere enclosing a point defect to an order parameter manifold $M$, it cannot continuously shrink to a point in the manifold. Thus, $g$ represents a nontrivial monopole. We can apply this procedure to the second Abe homotopy group. The second Abe homotopy group $\kappa_2 (M)$ consists of a homotopy class of a cylinder $S^1 \times [0,1]$, where both edges of a cylinder $S^1 \times \{0 \} \cup S^1 \times \{1 \}$ are mapped to the base point in the manifold. Thus, the cylinder in the manifold becomes a pinched torus as illustrated in Fig. \ref{letter-1} (a). When the tube shrinks to a line, then it changes to a loop. This map corresponds to an element of $\pi_1 (M)$. Similarly, when we remove the hole from the pinched torus, it becomes a sphere. Thus, $\kappa_2 (M)$ involves an element of $\pi_2 (M)$, too.         

Because a pinched torus can be continuously deformed to either a loop or a sphere, $\kappa_2 (M)$ includes both $\pi_1 (M)$ and $\pi_2 (M)$ as its subgroups. We apply this composite homotopy group to the case in which vortices and monopoles coexist. From Ref. \cite{Abe;1940}, elements of $\kappa_2 (M)$ are described by both $\pi_1 (M)$ and $\pi_2 (M)$, and the algebraic structure is represented by a semidirect product of $\pi_1 (M)$ and $\pi_2 (M)$ because $\pi_2 (M)$ does not always commute with $\pi_1 (M)$. This semidirect product can be interpreted as the influence of a vortex on monopoles. If there is no nontrivial influence on monopoles, $\kappa_2 (M)$ becomes the direct product group of $\pi_1 (M)$ and $\pi_2 (M)$.  Thus, we can decide whether or not there is a nontrivial influence by calculating the  commutator subgroup of $\pi_1(M)$ and $\pi_2(M)$. For given $\gamma \in \pi_1 (M)$ and $\alpha \in \pi_2 (M)$, we define the commutator $[\gamma ,\alpha] \equiv \gamma \ast \alpha \ast \gamma^{-1} \ast \alpha^{-1}$.  

To compute the commutator $[\gamma ,\alpha]$, we use the Eilenberg theorem \cite{Eilenberg;1939} which allows us to calculate an action of $\pi_1 (M)$ on $\pi_n (M)$. We define a mapping degree of freedom $\deg (f)$ as an integer coefficient of the $n$-th homology group in terms of an $n$-sphere $S^n$. Using the Eilenberg theorem for the action of $\pi_1 (M)$ on $\pi_n (M)$, the commutation relation gives
\begin{equation}
 \gamma (\alpha) = \alpha^{\deg(f_{\gamma})},
\end{equation}   
where $\gamma (\alpha) \equiv \gamma \ast \alpha \ast \gamma^{-1}$ and $f_{\gamma}$ is an action of $\gamma \in \pi_1 (M)$ on $S^n$. By considering the action $\gamma$ on $S^n$, we obtain $\deg (f_{\gamma}) = \pm 1$. When $\deg (f_{\gamma}) = +1$, hence $[\gamma ,\alpha] =1$, there is no influnce on monopoles. When $\deg(f) =-1$, there is an influence of $\pi_1 (M)$ on $\pi_2 (M)$. In particular, if $M=S^2 /\mathbb{Z}_2$, we can show that $\gamma (\alpha) = \alpha^{-1}$ because $f_\gamma$ corresponds to an antipodal map $f_{\gamma}(x) = -x$ with $x \in S^2$. Since the nontrivial element of $\pi_1 (S^2 /\mathbb{Z}_2 )$ describes an Alice string in our system, the monopole is influenced by the Alice string. Therefore,  $\kappa_2 (S^2/\mathbb{Z}_2)$ is isomorphic to $\mathbb{Z}_2 \ltimes \mathbb{Z}$, where $\mathbb{Z}_2$ represents Alice strings, and $\mathbb{Z}$ represents  monopoles,        
 where the algebraic structure of $\kappa_2 (S^2 /\mathbb{Z}_2)$ is described by using a semidirect product because $\pi_2 (S^2/\mathbb{Z}_2)$ does not commute with $\pi_1 (S^2 /\mathbb{Z}_2)$. Let us describe an element of $\kappa_2 (S^2 /\mathbb{Z}_2)$ as $(n,m)$, where the number of Alice strings is $n$ and the number of monopoles is $m$. As for the influence of Alice strings, 
the state $(1,1)$, which involves one Alice string and one monopole, can be annihilated by the presence of another $(1,1)$ because monopole charge $+1$ can be transformed to charge $-1$ if it makes a full circuit around the Alice string. This makes a sharp contrast with an ordinary situation in which adding of $(1,1)$ to $(1,1)$ gives $(2,2)$. Therefore, for $M= S^2/\mathbb{Z}_2$, there is a limit on the number of monopoles.      
\begin{figure}[htbp]
 \centering
 \includegraphics[width=8cm]{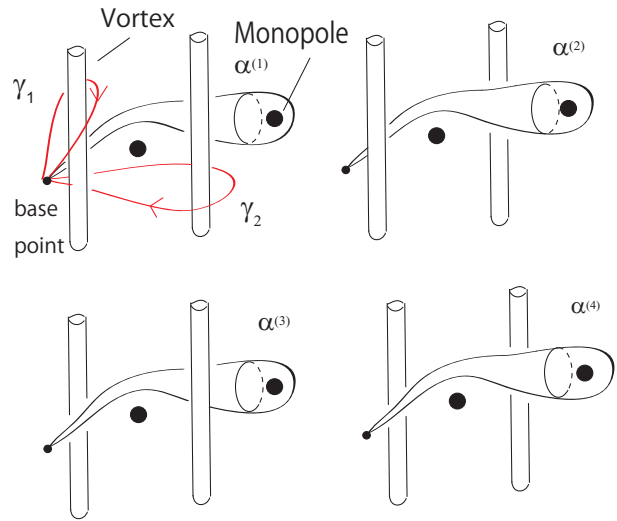}
 \caption{(Color online) Four configurations that do not continuously transform to each other, where a monopole between the vortice is circumvented. We label each configurations as $\alpha^{(1)},\alpha^{(2)},\alpha^{(3)}$ and $\alpha^{(4)}$. Both $\gamma_1$ and $\gamma_2$ represent indices of vortices.  }\label{letter-3}
\end{figure}

When there are two vortices and two monopoles as shown in Fig. \ref{letter-3}, we must consider four different configurations that do not shrink continuously to a point. Because the monopole charge depends crucially on the location of the base point, the following six relations hold corresponding to these configurations: 
\begin{equation}
\begin{split}
& \gamma_1^{-1} \ast \alpha^{(1)} \ast \gamma_1 = \alpha^{(3)}, \\
& \gamma_2^{-1} \ast \alpha^{(1)} \ast \gamma_2 = \alpha^{(2)}, \\
& \gamma_1 \ast \alpha^{(4)} \ast \gamma_1^{-1} = \alpha^{(2)}, \\
& \gamma_2 \ast \alpha^{(4)} \ast \gamma_2^{-1}  = \alpha^{(3)}, \\
& \gamma_2 \ast \gamma_1 \ast \alpha^{(4)} \ast \gamma_1^{-1} \ast \gamma_2^{-1} = \alpha^{(1)}, \\
& \gamma_2^{-1} \ast \gamma_1 \ast \alpha^{(3)} \gamma_1^{-1} \ast \gamma_2 = \alpha^{(2)}, \label{abe-phys-1}
\end{split}
\end{equation}
where $\gamma_1 $ and $\gamma_2 $ denote loops encircling the vortex near the base point and the next nearest neighbor, respectively. However, the path dependence of monopoles has no physical significance in a real situation. To eliminate the path dependence, we apply equivalent relations $\alpha^{(1)} \sim \alpha^{(2) } \sim \alpha^{(3)} \sim \alpha^{(4)}$ among the elements of $\kappa_2 (M)$. These relations give additional constrains on monopoles. Namely, it is necessarily for monopoles to satisfy     
\begin{equation}
 \gamma_i^{-1} \ast \alpha \ast \gamma_i \sim \alpha, \label{abe-2}
\end{equation}
where $i=1,2$. Equivalent relation (\ref{abe-2}) means that we have to take a conjugacy class of $\pi_1 (M)$ on an element of $\pi_2 (M)$. Similarly, we have to take a conjugacy class for $\pi_1 (M)$: 
\begin{equation}
 \gamma_1^{-1} \ast \gamma_2 \ast \gamma_1 \sim \gamma_2, \label{abe-3}
\end{equation}
where $\gamma_1 \neq \gamma_2 \in \pi_1 (M)$. Equivalent relation (\ref{abe-3}) means that an entanglement of vortices is independent of which of the base points we choose; thus the entanglement is invariant if we see from a different point of view \cite{Mermin;1979,Trebin;1982,Michel;1980,Mineev;1998,Volovic;2003}. Similarly, equivalent relation (\ref{abe-2}) represents that an entanglement between monopoles and vortices is independent of the base point.     
For the case of $M = S^2 /\mathbb{Z}_2$, by using equivalent relation(\ref{abe-2}) and $\gamma (\alpha) = \alpha^{-1} $ that describes an influence of an Alice string on a monople, we find that
 \begin{equation}
  \alpha^{2} \sim 1. \label{abe-4}
 \end{equation}
 We can denote an element of $\pi_2 (S^2 /\mathbb{Z}_2)$ by $\alpha^n $ with $ n \in \mathbb{Z}$. Thus, equivalent relation (\ref{abe-4}) reduces to $n \mod 2$. We do not only apply equivalent relation (\ref{abe-4}) to the case in the  presence of Alice strings, but also to the case in which there is no Alice string. A pair of Alice strings occur naturally as 
\begin{equation}
 (1 , \alpha ) \ast (1 , \alpha') \sim (\gamma_g ,\alpha ) \ast (\gamma_g , \alpha'),
\end{equation}
where $\alpha , \alpha' \in \pi_2 (S^2 /\mathbb{Z}_2), 1 , \gamma_g \in \pi_1 (S^2 /\mathbb{Z}_2)\ \ (\gamma_g^2 =1)$ \cite{footnote-3}. 
 Therefore, monopoles are affected due to the presence of an Alice string whenever it is possible to create a pair of Alice strings. Thus, we find that the monopole charge is isomorphic to $\mathbb{Z}_2$ for every element of $\pi_1 (S^2/\mathbb{Z}_2)$. Therefore, the second Abe homotopy group modulo the equivalent relation becomes 
\begin{equation}
 \kappa_2 (S^2 / \mathbb{Z}_2)/ \sim \ \  \cong \mathbb{Z}_2 \times \mathbb{Z}_2.
 \end{equation}

Finally, we point out that it is possible to create an entanglement between higher dimensional topological defects. Given $\alpha \in \pi_n (M)$ and $\beta \in \pi_m (M)$, a generalized commutator is defined by $[\alpha , \beta] \in \pi_{n+m-1} (M)$. The generalized commutator is isomorphic to the Whitehead product $\alpha \circ \beta$ \cite{J.Whitehead;1941,G.Whitehead;1978}. One of the nontrivial elements of a higher entanglement is $\pi_2 \circ \pi_2 \subset \pi_3 (M)$. This element represents a nontrivial composite soliton in three-dimensions or a nontrivial instanton in $2 +1$ dimensions. 

Higher dimentional Abe homotopy groups can describe an influence of $\pi_1 (M)$ on higher dimensional solitons. If $M =SO(3)$, $\pi_3 (SO(3)) \cong \mathbb{Z}$, there are $3$ dimensional skyrmions \cite{Shankar;1977,Volovik;1977}. Since $\pi_1 (SO(3)) \cong \mathbb{Z}_2$, the third Abe homotopy group $\kappa_3 (SO(3))$ is isomorphic to $\mathbb{Z}_2 \ltimes \mathbb{Z}$. However, the skyrmion charge does not change after rotating around a vortex because $\deg(f_\gamma) =+1$ for $M= SO(3)$ \cite{Eilenberg;1939}. Hence, we obtain $\kappa_3 (SO(3)) \cong \mathbb{Z}_2 \times \mathbb{Z}$, which means that $3$ dimensional skyrmion is independent of vortices. More generally, there is no influence of $\pi_1 (M)$ for orbit space $S^3/G$ for any discrete group $G$.     
The detailed account of these findings will be reported elsewhere.      

In conclusion, we have introduced a classification of topological defects by using the second Abe homotopy group, which is isomorphic to a semidirect product of $\pi_1 (M)$ and $\pi_2 (M)$, where $\pi_1 (M)$ represents the winding number of vortices and $\pi_2 (M)$ represents the monopole charge. The semidirect product of $\pi_1 (M)$ and $\pi_2 (M)$ represents the influence of a vortex on a monopole. The consistent charges $\mathbb{Z}_2 \times \mathbb{Z}_2$ of the whole system has been obtained by taking a conjugacy class of the second Abe group. If the order parameter manifold can accomodate Alice strings, a monopole charge can change its sign by rotating around an Alice string; even when there are no Alice strings, monopoles will be influenced by Alice strings via spontaneous pair creation of Alice strings. Therefore, there will eventually be only one or zero monopole left in the Universe.   


This work was supported by KAKENHI (22340114, 22103005, 22740219, 22740265, 20740141), Global COE Program ``the Physical Sciences Frontier" and the photon Frontier Network Program, MEXT, Japan.


\begin{thebibliography}{999}
\bibitem{footnote-3}
 Note that the notation of $(\gamma ,\alpha)$ implies the element of $\pi_1 (M)$ and $\pi_2 (M)$ rather than integer.
\end{thebibliography}
\end{document}